\documentclass[twocolumn,prl,showpacs,preprintnumbers]{revtex4-1}
\usepackage{amssymb}
\usepackage{amsmath}
\usepackage{graphicx}
\usepackage{epsfig}
\usepackage{subfigure}
\usepackage{amsfonts}
\usepackage{CJK}
\usepackage{multirow}
\usepackage[colorlinks,linkcolor=blue,citecolor=blue]{hyperref}
\usepackage[english]{babel} 

\begin{document}

\title{Experimental demonstration of one-shot coherence distillation:\\
High-dimensional state conversions}
\author{Shao-Jie Xiong$^{1,2}$}
\author{Zhe Sun$^{1}$}
\email{sunzhe@hznu.edu.cn}
\author{Xiaofeng Li$^{1}$}
\author{Qi-Ping Su$^{1}$}
\author{Zhengjun Xi$^{3}$}
\author{Li Yu$^{1}$}
\author{Jin-Shuang Jin$^{1}$}
\author{Jin-Ming Liu$^{2}$}
\author{Franco Nori$^{4,5}$}
\email{fnori@riken.jp}
\author{Chui-Ping Yang$^{1}$}
\email{yangcp@hznu.edu.cn}

\begin{abstract}
We experimentally investigate problems of one-shot coherence distillation
[Regula, Fang, Wang, and Adesso, Phys. Rev. Lett. 121, 010401 (2018)]. Based
on a set of optical devices, we design a type of strictly incoherent
operation (SIO), which is applicable in high-dimensional cases and can be
applied to accomplish the transformations from higher-dimensional states to
lower-dimensional states. Furthermore, a relatively complete process of the one-shot
coherence distillation is experimentally demonstrated for three- and
four-dimensional input states. Experimental data reveal an interesting
result: higher coherence distillation rates (but defective) can be reached by
tolerating a larger error $\varepsilon $. Our finding paves a fresh way in
the experimental investigation of quantum coherence conversions through
various incoherent operations.
\end{abstract}

\affiliation{$^{1}$~Department of Physics, Hangzhou Normal
University, Hangzhou 310036, China}
\affiliation{$^{2}$~State Key Laboratory of Precision Spectroscopy,
Department of Physics, East China Normal University, Shanghai 200062, China}
\affiliation{$^{3}$~College of Computer Science, Shaanxi Normal University,
Xi’an 710062, China}
\affiliation{$^4$~Theoretical Quantum Physics Laboratory,
RIKEN Cluster for Pioneering Research, Wako-shi, Saitama 351-0198, Japan}
\affiliation{$^5$~Department of Physics, University of Michigan, Ann Arbor,
MI 48109-1040, USA} 

\maketitle
\date{\today }

\emph{Instruction.}--- Quantum coherence, exhibiting the fundamental feature
of quantum superposition, marks the departure of quantum physics from
classical physics. Recently, problems of quantum coherence have attracted
considerable attention because these are essential for quantum
foundations\thinspace \cite{Review coherence-1, Review coherence-2,
Theory-coherence-1,Theory-coherence-2,Theory-coherence-3,Theory-coherence-4,Theory-coherence-5,Theory-coherence-6,Theory-coherence-7,coherence-Nori-1,coherence-Nori-2,coherence-Nori-3}
and could also have practical applications in a wide variety of fields, such
as quantum cryptography, quantum simulations, thermodynamics, quantum
metrology, transport theory, and quantum biology \thinspace \cite{coherence-phy-1,coherence-phy-simulation1,coherence-phy-simulation2,coherence-phy-non Mar,coherence-phy-2,coherence-phy-3,coherence-phy-4,coherence-phy-5, coherence-phy-bio, coherence-phy-bio-2}.
In order to characterize quantum coherence in a mathematically rigorous
and physically meaningful framework, the resource theory of quantum
coherence was developed\thinspace \cite{Review coherence-1,MIO,resource
coherence-1,DIO, resource coherence-2,resource coherence-3,resource
coherence-4,resource
coherence-5,resource
coherence-6,resource
coherence-7}. In this setting, coherence is regarded as a quantum resource
which provides necessary cost in accomplishing useful tasks. The free
states, i.e., the incoherent states, are defined with diagonal density
matrices in terms of the reference basis. Following that, free operations
are incoherent operations that act unchangeably on the assemblage of all
incoherent states. Many different definitions of incoherent operations are
motivated by various physical and mathematical requirements, e.g., maximally
incoherent operations (MIO)\thinspace \cite{MIO}, incoherent operations
(IO)\thinspace \cite{resource coherence-1}, dephasing-covariant incoherent
operations (DIO)\thinspace \cite{DIO}, and strictly incoherent operations
(SIO)\thinspace \cite{resource coherence-2}. The relations between each of
these sets are nontrivial, i.e., $\text{SIO}\subseteq \text{IO}\subseteq
\text{MIO and}~\text{SIO}\subseteq \text{DIO}\subseteq \text{MIO}$.

One of the most significant aspects in the coherence resource theory is to
realize coherence conversions under incoherent operations. Many efforts have
been devoted to explore the conditions for coherence manipulation and state
transformation\thinspace \cite{IOtransformation1,IOtransformation2, IO-Xi ZJ}%
. Quite recently, the experimental research on these problems was
reported\thinspace \cite{Exp-state-conversion}. The authors realized a kind
of SIO in two-dimensional (2D) space and implemented the state conversion on
qubits both with and without assistance.

A particularly important coherence conversion process is the so-called
coherence distillation, which focuses on the interconversion between copies
of a given state $\rho $ and a canonical unit resource $\left\vert \Phi
_{m}\right\rangle $ ($m$-dimensional maximally coherent state). The
asymptotic version was introduced in\thinspace \cite{resource coherence-3},
the authors used infinite copies of $\rho $ and $\left\vert \Phi
_{m}\right\rangle $ to check the asymptotical reversibility. The assisted
distillation proposal was presented in\thinspace \cite{Assisted distillation}%
, which is based on the interconversion between quantum correlation and
quantum coherence. The assisted coherence distillation was experimentally
tested in \thinspace \cite{Experimental assisted distillation}.

\begin{figure*}[tbp]
\begin{center}
\epsfig{file=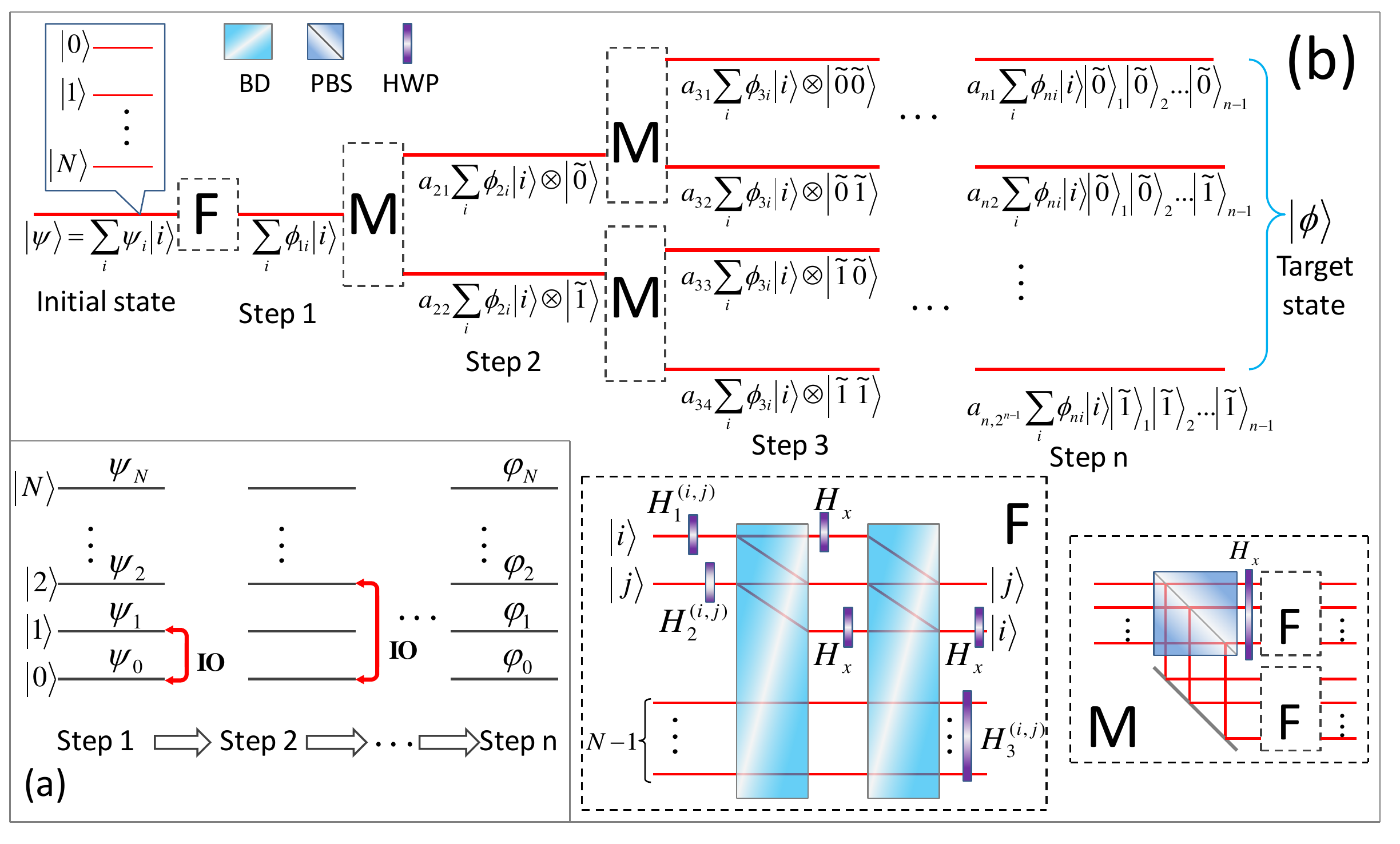,width=17.3cm, height=8.4cm}
\end{center}
\caption{ Experimental setup for realizing the multi-step proposal of the
SIO, which implements the pure state transformation $|\protect\psi \rangle
=\sum_{i=0}^{N}\protect\psi _{i}|i\rangle \rightarrow |\protect\phi \rangle
=\sum_{i=0}^{N^{\prime }}\protect\phi _{i}|i\rangle $, where $|i\rangle $
denotes the spatial modes and the dimensionality $N^{\prime }\leq N$. The
devices in the setup are half-wave plates (HWP), beam displacers (BD), and
polarizing beam splitters (PBS). (a) Sketch of the multi-step proposal. At
each step, only two components are manipulated. (b) Experimental setup of
the proposal, where the module \textbf{F} presents the two-dimensional SIO
working on the $|i\rangle $ and $|j\rangle $ spatial modes, holding the
other ($N-1$) modes unchanged. The angles of the HWP $H_{1}^{ij}$ and $%
H_{2}^{ij}$ are adjusted as needed. All the angles of the HWP $H_{x}$ are
set to $\protect\pi /4$. The angle of $H_{3}^{ij}$ is adjusted according to
the outputs of $i,j$ paths. In the module \textbf{M}, a PBS is employed to
destroy the quantum coherence appearing in the ancillary modes. A pair of
binary numbers $\widetilde{0}$ and $\widetilde{1}$ label the split path
groups. A set of binary numbers encodes an ancillary mode corresponding to a
group of spatial modes which carry the superposition information of the
target state. The superposition coefficients $\protect\phi _{ni}$ can be
converted to the coefficients $\protect\phi _{i}$ involved in the target
state. In addition, $a_{n1}$, $a_{n2}$, \ldots, and $a_{n2^{n-1}}$ satisfy $%
\sum_{k}^{2^{n-1}}a_{nk}^{2}=1$. }
\label{fig:1}
\end{figure*}

Note that the asymptotic distillation proposal lies on the assumed access to
an unbounded number of independently and identically distributed copies of
the considered system. In a realistic setting, only a finite supply of
states are available. Moreover, it is a huge challenge to collectively
manipulate coherent states over a large number of systems. Therefore, it is
necessary to consider a more general scenario, i.e., the one-shot version of
coherence distillation\thinspace \cite{One-Shot Distillation-1, One-Shot
Distillation-2, One-Shot Distillation-3, One-Shot Distillation-4}. In
particular, in the literature\thinspace \cite{One-Shot Distillation-1}, the
authors introduced a nonasymptotic coherence distillation, which requires a
single copy of a quantum system and adopts an $\varepsilon $-error fidelity
to characterize the distillation rate, achievable under a given class of
incoherent operations with an error tolerance $\varepsilon $. This scenario
greatly facilitates the experimental investigation.

Compared to the rapid development of theoretical work, there is a lack of
experimental investigation on the problems of the quantum coherence
conversion and the realization of different classes of incoherent
operations. Especially for high-dimensional cases, there are few reports of
the relevant experimental study. Whereas, realizing incoherent operations in
high-dimensional systems is a crucial subject in the quantum coherence
resource theory. For example, in coherence distillation, high-dimensional
incoherent operations are indispensable to convert the higher-dimensional
states into the lower-dimensional maximally coherent states. Motivated by
the above, here we propose a linear optical setup to experimentally realize
a type of incoherent operations which can be generalized to an arbitrarily
high-dimensional space. As an example, three- and four-dimensional (3D and
4D) cases are taken into account in our experiment. Based on the proposed
incoherent operation, we demonstrate the scenario of the one-shot coherence
distillation experimentally\thinspace \cite{One-Shot Distillation-1}. To the
best of our knowledge, this is the first experimental report on the one-shot
coherence distillation via converting higher-dimensional states into
lower-dimensional states. This proposal and observation might play important
roles in the future study on quantum coherence manipulation in physical
systems.

\begin{figure*}[tbp]
\begin{center}
\epsfig{file=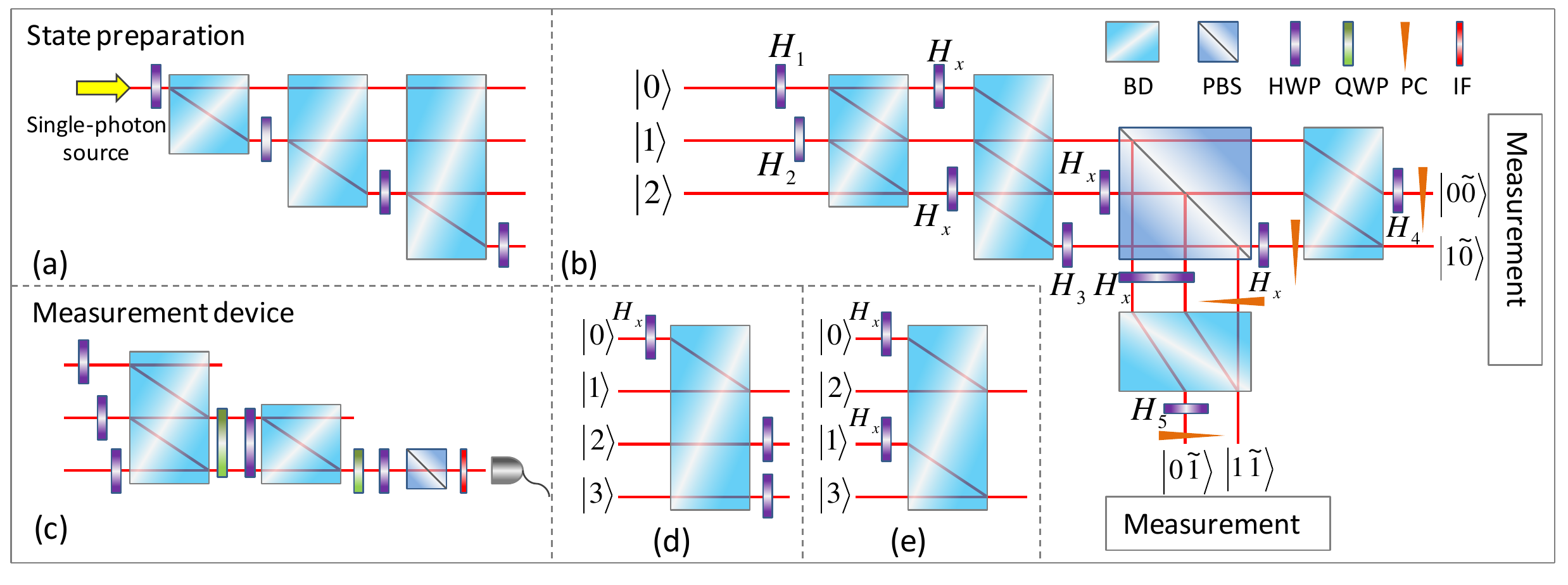,width=17.5cm, height=7cm}
\end{center}
\caption{Simplified setup for the SIO converting the three- and
four-dimensional pure states into the target states. (a) Device for
preparing pure states with no higher than four dimensions. (b) Setup for the
SIO converting a 3D pure state into a 2D pure state. The angles of the HWPs $H_{1,2,3,4,5}$ are adjusted as needed. All the angles of $H_{x}$ are set to
$\protect\pi /4$, which perform the inversions between the polarization modes, i.e., $|H\rangle\rightarrow|V\rangle$  and  $|V\rangle\rightarrow|H\rangle$.
(c) Setup for the spatial tomography measurement on 3D states. (d) and (e) show the operations combining two paths into one path,
which accomplish the dimension reductions. Some other devices in the setup
are quarter-wave plates (QWP), phase compensators (PC), and interference filters (IF).}
\label{fig:2}
\end{figure*}

\emph{One-shot coherence distillation.---} The concept of asymptotic
coherence distillation was first put forward in\thinspace \cite{resource
coherence-3}. In the asymptotic limit, an unbounded number of the state
copies are needed, which is however quite difficult to achieve in a
realistic setting. To overcome this difficulty, Regula and
co-workers\thinspace \cite{One-Shot Distillation-1} introduced the one-shot
coherence distillation tolerating an error $\varepsilon $, which is measured
by%
\begin{equation}
C_{d,\mathcal{O}}^{(1),\varepsilon }(\rho ):=\log \max \{m\in \mathbb{N}|F_{%
\mathcal{O}}(\rho ,\left\vert \Phi _{m}\right\rangle )\geq 1-\varepsilon \},
\label{Cd_one-shot}
\end{equation}%
where the superscript \textquotedblleft $(1)$\textquotedblright\ indicates
that only a single copy of the given state $\rho $ (or pure state $|\psi
\rangle $) and the $m$-dimensional maximally coherent state $\left\vert \Phi
_{m}\right\rangle $ are included. The asymptotic version is obtained in the
limit $C_{d,\mathcal{O}}^{\infty ,\varepsilon }(\rho )=\underset{\varepsilon
\rightarrow 0}{\lim^{{}}}\underset{n\rightarrow 0}{\lim }C_{d,\mathcal{O}%
}^{(1),\varepsilon }(\rho ^{\otimes n})/n$. The definition of the
distillation fidelity $F_{\mathcal{O}}(\rho ,\left\vert \Phi
_{m}\right\rangle )$ is
\begin{equation}
F_{\mathcal{O}}(\rho ,\left\vert \Phi _{m}\right\rangle )=\max_{\Lambda \in
\mathcal{O}}\langle \Lambda (\rho ),\left\vert \Phi _{m}\right\rangle
\rangle ,  \label{f_distillation}
\end{equation}%
where $\mathcal{O}$ denotes a class of incoherent operations and $\langle
A,B\rangle =\text{Tr}(A^{\dagger }B)$ is the Hilbert-Schmidt inner product.
This distillation fidelity describes the maximal conversion rate from a
given state to the maximally coherent state $\left\vert \Phi
_{m}\right\rangle $ by optimizing the incoherent operations. The value of $%
F_{\mathcal{O}}(\rho ,\left\vert \Phi _{m}\right\rangle )$ depends on the
structure of the given state, the chosen dimension $m$ of $\left\vert \Phi
_{m}\right\rangle $, and the type of the incoherent operations. The error $%
\varepsilon $ in Eq.\thinspace (\ref{Cd_one-shot}) introduces rich meanings,
such as the \textquotedblleft purity\textquotedblright\ or \textquotedblleft
quality\textquotedblright\ to the distillable coherence. In practice, by
tolerating a larger error $\varepsilon $ (i.e., more defects), one can
obtain a higher distillation rate (i.e., larger $m$).

The key step in the one-shot coherence distillation is to realize proper
incoherent operations. We will design a linear optical device to
experimentally realize the incoherent operation applicable for
high-dimensional cases. In this letter, only pure-state conversions are
studied for the following reasons: (i) Pure states are important resource in
quantum tasks; (ii) Theoretical results are clear for pure states, e.g., the
one-shot distillable coherence of pure states under MIO, DIO, IO, or\ SIO is
exactly the same\thinspace \cite{One-Shot Distillation-1}, and (iii) We will
propose a SIO to accomplish the one-shot coherence distillation in pure
states.

\emph{Pure state transformations under incoherent operations}.--- Let us
start with a 2D SIO. In realistic settings, incoherent operations in a
primary system are usually performed by introducing ancillary systems. In
the linear optical setup, we employ the spatial modes of the photons, i.e., $%
\left\vert 0\right\rangle $ and $\left\vert 1\right\rangle $, to describe\
the primary system state, which facilitates the expansion to
high-dimensional cases. The polarization modes $\left\vert V\right\rangle $
and $\left\vert H\right\rangle $ (i.e., vertical and horizontal modes) act
as the ancillary system. The experimental setup is shown in the module
\textbf{F }of Fig.\thinspace 1, where the part of the $\left\vert
i\right\rangle $ and $\left\vert j\right\rangle $ modes denotes the 2D SIO.
Here, we select the modes $\left\vert 0\right\rangle $ and $\left\vert
1\right\rangle $, for example. We input a product state of the total system,
i.e., $\left\vert \psi \right\rangle \left\vert V\right\rangle =\left(
\alpha \left\vert 0\right\rangle +\beta \left\vert 1\right\rangle \right)
\left\vert V\right\rangle $, with real numbers $\alpha $ and $\beta $. The
angles $\theta _{1}$ and $\theta _{2}$ of the half-wave plates (HWP) $%
H_{1,2}^{ij}$ are adjusted as needed. Then, the map $\Lambda (\left\vert
\psi \right\rangle \left\langle \psi \right\vert )=K_{1}\left\vert \psi
\right\rangle \left\langle \psi \right\vert K_{1}^{\dagger }+K_{2}\left\vert
\psi \right\rangle \left\langle \psi \right\vert K_{2}^{\dagger }$ can be
achieved with the Kraus operators
\begin{eqnarray}
K_{1} &=&\sin 2\theta _{1}|0\rangle \langle 0|+\cos 2\theta _{2}|1\rangle
\langle 1|,  \notag \\
K_{2} &=&\cos 2\theta _{1}|0\rangle \langle 1|+\sin 2\theta _{2}|1\rangle
\langle 0|.  \label{Kraus-qubit}
\end{eqnarray}%
When the parameters satisfy $\left\vert \alpha \right\vert ^{2}\sin (4\theta
_{1})=\left\vert \beta \right\vert ^{2}\sin (4\theta _{2})$, one can obtain
the pure output state, i.e., $\Lambda (\left\vert \psi \right\rangle
\left\langle \psi \right\vert )=\left\vert \phi \right\rangle \left\langle
\phi \right\vert $. According to the definition of SIO\thinspace \cite%
{IOtransformation1,IOtransformation2}, the operations described by the Kraus
operators in Eq.\thinspace (\ref{Kraus-qubit}) belong to a SIO.

For high-dimensional cases, we divide the operations into several steps. The
sketch of the proposal is shown in Fig.\thinspace 1(a), where the conversion
from the input pure state $\left\vert \psi \right\rangle $ to the target
pure state $\left\vert \phi \right\rangle $ can be realized by $n$ steps. At
each step, an incoherent operation (in fact, a SIO here) works on the two
components $\psi _{i_{{}}}$ and $\psi _{j}$. Since this elementary operation
belongs to a SIO, the following multi-step operation also belongs to a
SIO\thinspace \cite{supplement}.

In Fig.\thinspace 1(b), we present an experimental setup for realizing our
multi-step proposal. In the first step, a 2D SIO (see module \textbf{F)}
works on the two components $|i\rangle $ and $|j\rangle $, holding the other
components unchanged. After that, module \textbf{M }is introduced, where a
polarization beam splitter (PBS) is employed to reset the superposed
ancillary modes into a single one. Then, module \textbf{F }is repeatedly
applied to manipulate another pair of spatial modes different from the pair
in the previous step. At the {n-th\ step, there are $2^{n-1^{{}}}$ copies of
the target state corresponding to $2^{n-1^{{}}}$ groups of spatial modes.
One can obtain the target state, deterministically by performing spatial
tomography on all the outputs, but probabilistically by reading part of the
outputs. We implement the deterministic detection to complete the one-shot
coherence distillation. }

\emph{Experimental demonstration of the one-shot coherence distillation.}---
A single-photon source is produced by pumping a type I $\beta $-barium
borate crystal with ultraviolet pulses at a 405-nm centered wavelength. One
photon is directly detected as a trigger. The other one is prepared in a
pure state of the spatial modes $\left\vert i\right\rangle $ ($i=0,1,2,...$%
). Figure\thinspace 2(a) shows the state preparation. A 3D pure state and a
4D pure state, with one undetermined superposition coefficient, are chosen
as the input states.

\emph{Three-dimensional distillation.}--- The input state is chosen as
\begin{equation}
|\psi ^{3}\rangle =\sqrt{\alpha }|2\rangle +\sqrt{(1-\alpha )/2}(|0\rangle
+|1\rangle ),  \label{3D-input}
\end{equation}%
where $~\alpha \in\lbrack 0,1]$. The superscript \textquotedblleft $3$%
\textquotedblright\ denotes the dimensionality. Based on the distillation
fidelity in Eq.\thinspace (\ref{f_distillation}), one should obtain the
target states closest to the maximally coherent states by performing proper
incoherent operations. In the region $\alpha \in \left[ 0,1/2\right] $, the
distillation fidelity $F_{\mathcal{O}}(|\psi ^{3}\rangle ,\left\vert \Phi
_{2}\right\rangle )$ is proved to be $1\,$\cite{supplement}. Theoretically,\
$|\psi ^{3}\rangle $ can be perfectly converted to the maximally coherent
state, $\left\vert \Phi _{2}\right\rangle =(|0\rangle +|1\rangle )/\sqrt{2}$.

Figure\thinspace 2(b) presents an experimental setup to accomplish the
conversion from $|\psi ^{3}\rangle $ to $\left\vert \Phi _{2}\right\rangle $%
. It is a simplified two-step version of the general proposal in
Fig.\thinspace 1(b). After the PBS, only one beam displacer (BD) is needed
to combine two paths into one, resulting in a dimension reduction. By
adjusting the parameters $\theta _{1,2,3,4,5}$, one obtains $\left\vert \Phi
_{2}\right\rangle $ by doing a spatial tomography [see Fig.\thinspace 2(c)].
In fact, the proposal in Fig.\thinspace 2(b) provides a general SIO which
can convert 3D pure states to 2D pure states. The parameterized Kraus
operators are shown in~\cite{supplement}.

In the region $\alpha \in (1/2,1]$, the distillation fidelity becomes $F_{%
\mathcal{O}}(|\psi ^{3}\rangle ,\left\vert \Phi _{2}\right\rangle )=\frac{1}{%
2}(\sqrt{\alpha }+\sqrt{1-\alpha })^{2}$ (see derivations in\thinspace \cite%
{supplement}), which implies that the maximally coherent state $\left\vert
\Phi _{2}\right\rangle $ cannot be reached. Instead, a possible target state
is $|\phi _{3\rightarrow 2}\rangle =\sqrt{\alpha }|0\rangle +\sqrt{1-\alpha }%
|1\rangle $. We design a particularly\ simple device in Fig.\thinspace 2(d),
where three input paths ($|0\rangle $, $|1\rangle $, $|2\rangle $) are
enough to accomplish the conversion $|\psi ^{3}\rangle \rightarrow |\phi
_{3\rightarrow 2}\rangle $. A beam displacer is employed to combine the
spatial modes $|0\rangle $ and $|1\rangle $. The angle of the HWP in the
path of $|2\rangle $ is adjusted according to the initial superposition
coefficients of $|0\rangle $ and $|1\rangle $. All the operations, performed
on the spatial modes, belong to SIO, which can be verified by their
Kraus-operator representation\thinspace \cite{supplement}.

Figure\thinspace 3(a) demonstrates the experimental data of $F_{\mathcal{O}%
}(|\psi ^{3}\rangle ,\left\vert \Phi _{2}\right\rangle )$ and $F_{\mathcal{O}%
}(|\psi ^{3}\rangle ,\left\vert \Phi _{3}\right\rangle )$ versus the
superposition coefficient $\alpha $. Here, the state $\left\vert \Phi
_{3}\right\rangle $ is a 3D maximally coherent state, i.e., $\left\vert \Phi
_{3}\right\rangle =(|0\rangle +|1\rangle +|2\rangle )/\sqrt{3}$.\ To obtain
the fidelity $F_{\mathcal{O}}(|\psi ^{3}\rangle ,\left\vert \Phi
_{3}\right\rangle )$, we implement no operation except tomography
measurements at the outputs. The experimental data are denoted by the
triangules. To the fidelity $F_{\mathcal{O}}(|\psi ^{3}\rangle ,\left\vert
\Phi _{2}\right\rangle )$, the corresponding SIO is realized by the devices
shown in Fig.\thinspace 2(b, d). The dependence of the parameters $\theta
_{1,2,3,4,5}$ [in Fig.\thinspace 2(b)] on the coefficient $\alpha $ is shown
in the Supplemental Material~\cite{supplement}.\ The experimental data for $%
F_{\mathcal{O}}(|\psi ^{3}\rangle ,\left\vert \Phi _{2}\right\rangle )$,
denoted by the rhombuses, agree well with the theoretical curve. In the
region $\alpha \in \lbrack 0,1/2]$, $F_{\mathcal{O}}(|\psi ^{3}\rangle
,\left\vert \Phi _{2}\right\rangle )$ approaches to $1$; while in the region
$\alpha \in \lbrack 1/2,1]$, $F_{\mathcal{O}}(|\psi ^{3}\rangle ,\left\vert
\Phi _{2}\right\rangle )$ decreases to $1/2$ for increasing values of $%
\alpha $.
\begin{figure}[tbp]
\begin{center}
\epsfig{file=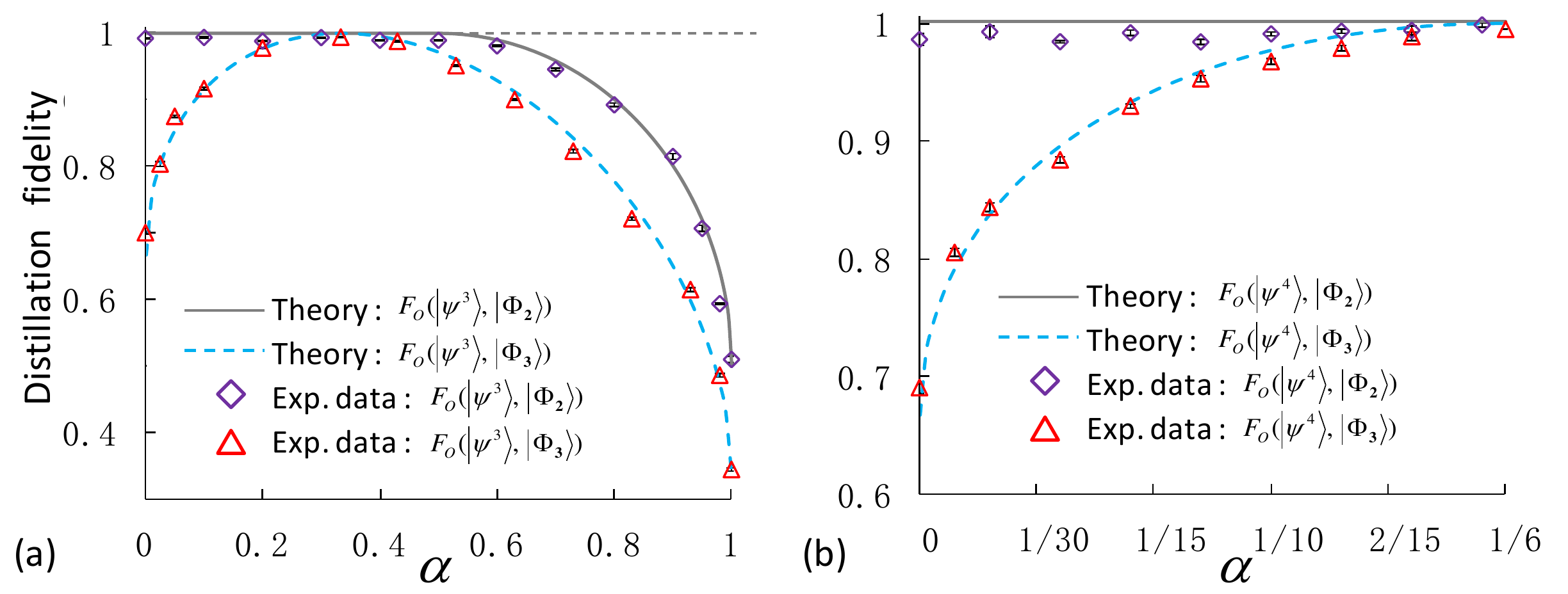,width=9.35cm, height=4.4cm}
\end{center}
\caption{Experimental data of the distillation fidelity vs the superposition
coefficient $\protect\alpha $. (a) Distillation fidelity of\ the 3D input
state $\left\vert \protect\psi ^{3}\right\rangle $ given in Eq.\thinspace (%
\protect\ref{3D-input}). (b) Distillation fidelity of the 4D input state $%
\left\vert \protect\psi ^{4}\right\rangle $ given in Eq.\thinspace (\protect
\ref{4D-input}). $\left\vert \Phi _{2}\right\rangle $ and $\left\vert \Phi
_{3}\right\rangle $ are the maximally coherent states of 2D and 3D,
respectively. }
\label{fig:3}
\end{figure}
According to Eq.\thinspace (\ref{Cd_one-shot}), one can see that if a zero
error $\varepsilon =0$ is strictly required, the distillable coherence will
be measured as $C_{d,\mathcal{O}}^{(1),\varepsilon =0}(|\psi ^{3}\rangle
)=\log 2$ in the region $\alpha \in \lbrack 0,1/2]$, except at the point $%
\alpha =1/3$ where $C_{d,\mathcal{O}}^{(1),\varepsilon =0}(|\psi ^{3}\rangle
)=\log 3$. However, when $\alpha >1/2$, the distillation\ fidelity cannot
reach $1$ any more, i.e., $C_{d}^{(1),\varepsilon =0}(|\psi ^{3}\rangle )=0$%
. This result implies that the ideal coherence resource cannot be distilled
from the initial state. However, in practical tasks, a finite tolerance $%
\varepsilon \neq 0$ is usually accepted, then one can finish the tasks with
a defective distillable coherence in larger regions of $\alpha $. For
example, if an accepted error is$\ \varepsilon =0.1$, the distillable
coherence will be $C_{d,\mathcal{O}}^{(1),\varepsilon =0.1}(|\psi
^{3}\rangle )=\log 3$ in a larger region about$\ \alpha \in \lbrack
0.0838,0.6495]$, and $C_{d,\mathcal{O}}^{(1),\varepsilon =0.1}(|\psi
^{3}\rangle )=\log 2$ in $\alpha \in \lbrack 0,0.0838)\cup (0.6495,0.8]$.
Such an example clearly presents the fact that in the one-shot distillation
framework, one can obtain a higher rate of distillable coherence but with
more defects by tolerating a larger error $\varepsilon $. Similar phenomena
are also found in Fig.\thinspace 3(b) where a 4D input state is investigated.

\emph{Four-dimensional distillation.}--- We choose a 4D input state (for $%
\alpha \in \lbrack 0,1/2]$)
\begin{equation}
|\psi ^{4}\rangle =\sqrt{\alpha }(|0\rangle +|1\rangle )+\sqrt{1/2-\alpha }%
(|2\rangle +|3\rangle ).  \label{4D-input}
\end{equation}

For the 2D maximally coherent state $\left\vert \Phi _{2}\right\rangle $,
the distillation fidelity $F_{\mathcal{O}}(|\psi ^{4}\rangle ,\left\vert
\Phi _{2}\right\rangle )$ is proved to be $1$\thinspace \cite{supplement}
over the entire range $\alpha \in \lbrack 0,1/2]$. Thus a reasonable target
state is $|\Phi _{2}\rangle =(|2\rangle +|3\rangle )/\sqrt{2}$. The general
proposal shown in Fig.\thinspace 1(b) can be employed to reach the target
state. However, due to the special structure of $|\psi ^{4}\rangle $, a
simplified device in Fig.\thinspace 2(e) is designed by using a BD to
combine the paths $|0\rangle $ and $|1\rangle $ into the paths $|2\rangle $
and $|3\rangle $, respectively. This finishes the state conversion.

For the 3D maximally coherent state $\left\vert \Phi _{3}\right\rangle $,
the distillation fidelity is $F_{\mathcal{O}}(|\psi ^{4}\rangle ,\left\vert
\Phi _{3}\right\rangle )=\frac{2}{3}(\sqrt{\alpha }+\sqrt{1-2\alpha })^{2}$
in the region of $\alpha \in \lbrack 0,1/6]\cup \lbrack 1/3,1/2]$\thinspace
\cite{supplement}. Therefore, a reasonable target state is $\left\vert \phi
_{4\rightarrow 3}\right\rangle =\sqrt{2\alpha }|1\rangle \!+\!\sqrt{%
(1/2-\alpha )}(|2\rangle +|3\rangle )$. While, in the region $\alpha \in
\lbrack 1/6,1/3]$, the fidelity is $F_{\mathcal{O}}(|\psi ^{4}\rangle
,\left\vert \Phi _{3}\right\rangle )=1$. Thus, the maximally coherent state $%
|\Phi _{3}\rangle $ becomes the target state .

The conversion $|\psi ^{4}\rangle \rightarrow \left\vert \phi _{4\rightarrow
3}\right\rangle $ can be accomplished through\ the devices in Fig.\thinspace
2(d) by combining the path $|0\rangle $ into $|1\rangle $. However, it is
much more complicated to realize the transformation $|\psi ^{4}\rangle
\rightarrow |\Phi _{3}\rangle $. It can be achieved either by the general
method (in Fig.\thinspace 1) with three steps, or by the simplified proposal
in Fig.\thinspace 2(b) with the extension to a four-path input and two more%
\textbf{\ }module \textbf{F }at the outputs.

In Fig.\thinspace 3(b), we show the experimental data of the distillation
fidelity $F_{\mathcal{O}}(|\psi ^{4}\rangle ,\left\vert \Phi
_{2}\right\rangle )$ (denoted by rhombuses)\ and $F_{\mathcal{O}}(|\psi
^{4}\rangle ,\left\vert \Phi _{3}\right\rangle )$ (denoted by triangles)
versus the superposition coefficient $\alpha $. In experiments, we test only
the region $\alpha \in \lbrack 0,1/6]$, where one can see that if the zero
error is strictly defined, the distillable coherence is measured by $%
C_{d}^{(1),\varepsilon =0}(|\psi ^{4}\rangle )=\log 2$, except at the point
of $\alpha =1/6$, where $C_{d}^{(1),\varepsilon =0}(|\psi ^{4}\rangle )=\log
3$. If a finite error $\varepsilon \neq 0$ is allowed, one will obtain a
higher rate $C_{d}^{(1),\varepsilon }(|\psi ^{4}\rangle )=\log 3$ in a wider
region of $\alpha $ conditioned by $F_{\mathcal{O}}(|\psi ^{4}\rangle
,\left\vert \Phi _{3}\right\rangle )\geq 1-\varepsilon $.

\emph{Discussion}.--- We have studied the problems of implementing
incoherent operations in a realistic optical system and demonstrated the
one-shot coherence distillation process experimentally. Based on an optical
setup, a general proposal was introduced to realize an important SIO
applicable in high dimensions, which accomplishes pure-state conversions.
Two sets of states were chosen as input state examples and their
distillation fidelities were obtained analytically. We clearly presented the
selection process of the target state and the experimental realization of
the incoherent operations. Therefore, this experiment provides a relatively
complete demonstration of the coherence distillation process. The
experimental data reveal that one can obtain a higher coherence distillation
rate (but with more defects) by tolerating a larger error $\varepsilon $. It
is hence of significance in practical tasks. Additional unitary operations
can be added onto the present setup to realize other kinds of incoherent
operations. Our finding opens a window through which one can explore in
depth the experimental implementation of quantum coherence conversions
through various incoherent operations.

This work is supported by the NKRDP of China (2016YFA0301802), the National
Natural Science Foundation of China (NSFC) (11375003, 11775065, 11774076,
11174081, 61472114,11974096,61671280), the Zhejiang Natural Science
Foundation (LY17A050003), and the Natural Science Foundation of Shanghai
(16ZR1448300). This work is partially supported by MURI Center for Dynamic
Magneto-Optics via the Air Force Office of Scientific Research (AFOSR)
(FA9550-14-1-0040), Army Research Office (ARO) (Grant No. Grant No.
W911NF-18-1-0358), Asian Office of Aerospace Research and Development
(AOARD) (Grant No. FA2386-18-1-4045), Japan Science and Technology Agency
(JST) (via the Q-LEAP program, and the CREST Grant No. JPMJCR1676), Japan
Society for the Promotion of Science (JSPS) (JSPS-RFBR Grant No.
17-52-50023; JSPS-FWO Grant No. VS.059.18N), the RIKEN-AIST Challenge
Research Fund, the Foundational Questions Institute (FQXi), and the NTT PHI
Laboratory.

\clearpage

\part{Supplemental Materials}

\section{Kraus representation of the proposed incoherent operations}

\begin{figure}[tbp]
\begin{center}
\epsfig{file=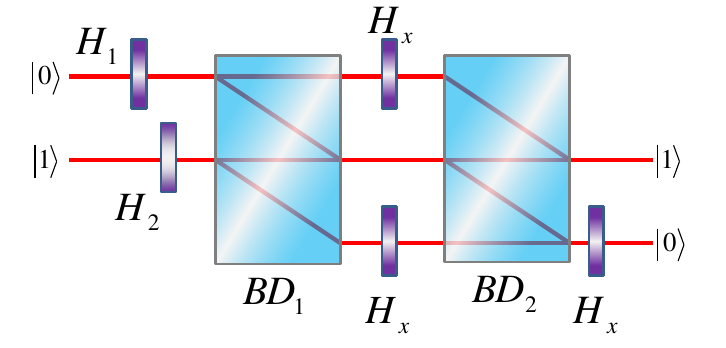,width=5.5cm, height=3cm}
\end{center}
\caption{Experimental setup for a 2D SIO, which realizes the transformations
between 2D pure states. $H_{1,2,x}$ denote the half-wave plates (HWP), and $%
\text{BD}_{1,2}$ denote the beam displacers (BD). The angle of ${H}_{x}$ is
set to $\protect\pi /4$, and the angles of ${H}_{1,2}$ are set according to
the target states.}
\label{fig:4}
\end{figure}

Let us introduce the proposed SIO in detail. Figure~\ref{fig:4} presents
more clearly the experimental setup of the 2D SIO. For the state
transformation in a 2D space $|\psi \rangle =\sum_{i=0}^{1}\psi
_{i}|i\rangle \rightarrow |\phi \rangle =\sum_{i=0}^{1}\phi _{i}|i\rangle $,
the Kraus-operator representation is $|\phi \rangle \left\langle \phi
\right\vert =\sum_{i=1}^{2}K_{i}|\psi \rangle \left\langle \psi \right\vert
K_{i}^{\dagger }$, and the Kraus operators can be described as\thinspace
\cite{IOtransformation1}
\begin{eqnarray}
K_{1} &=&\sqrt{a}\,\frac{\phi _{0}}{\psi _{0}}|0\rangle \langle 0|+\sqrt{a}\,%
\frac{\phi _{1}}{\psi _{1}}|1\rangle \langle 1|,  \notag \\
K_{2} &=&\sqrt{1-a}\,\frac{\phi _{0}}{\psi _{1}}|0\rangle \langle 1|+\sqrt{%
1-a}\,\frac{\phi _{1}}{\psi _{0}}|1\rangle \langle 0|,
\end{eqnarray}%
where displayed $a=\frac{\left\vert \psi _{0}\right\vert ^{2}-\left\vert
\phi _{1}\right\vert ^{2}}{\left\vert \phi _{0}\right\vert ^{2}-\left\vert
\phi _{1}\right\vert ^{2}}$ and $0\leq a\leq 1$, which is equivalent to
ensure the majorization relation. A pure state $\left\vert \psi
\right\rangle =\sum\nolimits_{i=0}^{N}\psi _{i}\left\vert i\right\rangle $,
majorized by another state $\left\vert \phi \right\rangle
=\sum\nolimits_{i=0}^{N}\phi _{i}\left\vert i\right\rangle $, should satisfy
$\sum_{i=0}^{k}|\psi _{i}|^{2\downarrow }\leq \sum_{i=0}^{k}|\phi
_{i}|^{2\downarrow }$, where\ $k\in \lbrack 0,N]$ and the superscript
\textquotedblleft $\downarrow $\textquotedblright\ indicates the descending
order of the elements. The majorization relation is sufficient and necessary
for SIO (or special IO)-dominated pure states conversions\thinspace \cite%
{One-Shot Distillation-3,IOtransformation2}.

The Kraus operators above can be rewritten as
\begin{eqnarray}
K_{1} &=&\sin 2\theta _{1}|0\rangle \langle 0|+\cos 2\theta _{2}|1\rangle
\langle 1|,  \notag \\
K_{2} &=&\cos 2\theta _{1}|0\rangle \langle 1|+\sin 2\theta _{2}|1\rangle
\langle 0|.  \label{2D-Kraus}
\end{eqnarray}%
Actually, in the experiment we realize a map
\begin{eqnarray}
|0V\rangle &\rightarrow &\cos (2\theta _{1})|1H\rangle +\sin (2\theta
_{1})|0V\rangle ,  \notag \\
|1V\rangle &\rightarrow &\cos (2\theta _{2})|1V\rangle +\sin (2\theta
_{2})|0H\rangle ,
\end{eqnarray}%
which can be translated into the Kraus-operator representation in
Eq.\thinspace (\ref{2D-Kraus}). For a pure input state $\left\vert \psi
\right\rangle =\alpha \left\vert 0\right\rangle +\beta \left\vert
1\right\rangle $, if the angles $\theta _{1,2}$ of the HWP (denoted by $%
H_{1,2}$ in Fig.~\ref{fig:4}) satisfy $\left\vert \alpha \right\vert
^{2}\sin (4\theta _{1})=\left\vert \beta \right\vert ^{2}\sin (4\theta _{2})$%
, one can obtain the pure state at the output, i.e.,
\begin{eqnarray}
\left\vert \phi \right\rangle &=&\frac{1}{Q}\left[ \beta \sin (2\theta
_{2})\left\vert 0\right\rangle +\alpha \cos (2\theta _{1})\left\vert
1\right\rangle \right] ,  \notag \\
Q &=&\sqrt{\left\vert \beta \right\vert ^{2}\sin ^{2}(2\theta
_{2})+\left\vert \alpha \right\vert ^{2}\cos ^{2}(2\theta _{1})}.
\end{eqnarray}
Recall the definition of IO and SIO. For a chosen reference basis $%
\{|i\rangle \}$, the class of free states is denoted by $\mathcal{I}$. IO
and SIO can be described by a set of Kraus operators $\left\{ K_{n}\right\} $
satisfying $\sum_{n}K_{n}^{\dagger }K_{n}=1$. For an IO, every Kraus
operator should satisfy $K_{n}\mathcal{I}K_{n}^{\dagger }\subseteq \mathcal{I%
}$. While, for a SIO, an additional condition, i.e., $K_{n}^{\dagger }%
\mathcal{I}K_{n}\subseteq \mathcal{I}$, is needed. An operation is IO if and
only if every column of $K_{n_{{}}}$ in the fixed basis $\{|i\rangle \}$ has
at most one nonzero entry. More strictly, SIO requires that not only every
column but also every line of $K_{n_{{}}}$ has at most one nonzero
element\thinspace \cite{IOtransformation1,IOtransformation2}. Therefore, the
2D operation proposed by us belongs to the SIO.

In order to consider the multi-step operations applicable to
high-dimensional space, we introduce the SIO performed on the subspace
spanned by the two modes $|i\rangle $ and $|j\rangle $, with its experiment
setup in the module \textbf{F} of Fig.\thinspace 1. The whole map, in the
composite system of the spatial modes and the ancillary polarization modes ($%
|H\rangle ^{{}}$and $|V\rangle $), reads
\begin{eqnarray}
&|iV\rangle \rightarrow &\cos 2\theta _{1}^{(i,j)}|jH\rangle +\sin 2\theta
_{1}^{(i,j)}|iV\rangle ,  \notag \\
&|jV\rangle \rightarrow &\cos 2\theta _{2}^{(i,j)}|jV\rangle +\sin 2\theta
_{2}^{(i,j)}|iH\rangle ,  \notag \\
&|kV\rangle \rightarrow &\left\vert k\right\rangle \left[ \cos 2\theta
_{3}^{(i,j)}|V\rangle +\sin 2\theta _{3}^{(i,j)}|H\rangle \right] ,
\end{eqnarray}%
where the parameters $\theta _{1}^{(i,j)}$ and $\theta _{2}^{(i,j)}$
correspond to the two HWPs $H_{1}^{ij}$ and $H_{2}^{ij}$ in the module
\textbf{F. }The\ mode $\left\vert k\right\rangle $ ($k\neq i$ and $j$)
denote the modes different from the $i$ and $j$ modes, and the angle $\theta
_{3}^{(i,j)}$ corresponding to the HWP $H_{3}^{ij}$ should be adjusted to
prepare the same superposition structure of the polarization modes as those
assisting the spatial modes of $|i\rangle $ and $|j\rangle $.

After this step, we realize the transformation
\begin{equation}
\sum_{i}\psi _{i}|i\rangle \rightarrow \phi _{i}|i\rangle +\phi
_{j}|j\rangle +\sum_{k}\psi _{k}|k\rangle .
\end{equation}
The Kraus operators can be expressed as
\begin{eqnarray}
K_{1}^{(i,j)}\! &=&\!\sin 2\theta _{1}^{(i,j)}|i\rangle \langle i|\!+\!\cos
2\theta _{2}^{(i,j)}|j\rangle \langle j|\!+\!\cos 2\theta _{3}^{(i,j)}I_{k},
\notag \\
K_{2}^{(i,j)}\! &=&\!\cos 2\theta _{1}^{(i,j)}|i\rangle \langle j|\!+\!\sin
2\theta _{2}^{(i,j)}|j\rangle \langle i|\!+\!\sin 2\theta _{3}^{(i,j)}I_{k},
\notag \\
I_{k} &=&\sum_{k\neq i,j}|k\rangle \langle k|,
\end{eqnarray}%
where the parameters
\begin{eqnarray}
2\theta _{1}^{(i,j)} &=&\arcsin (\sqrt{a_{ij}}\frac{\phi _{i}}{\psi _{i}}),
\notag \\
2\theta _{2}^{(i,j)} &=&\arccos (\sqrt{a_{ij}}\frac{\phi _{j}}{\psi _{j}}),
\notag \\
2\theta _{3}^{(i,j)} &=&\arcsin \sqrt{a_{ij}},
\end{eqnarray}
with $a_{ij}=\frac{\left\vert \psi _{i}\right\vert ^{2}-\left\vert \phi
_{j}\right\vert ^{2}}{\left\vert \phi _{i}\right\vert ^{2}-\left\vert \phi
_{j}\right\vert ^{2}}$. As a consequence, in the multi-step proposal, the
Kraus operators are implemented on different two-dimensional subspaces.\ One
of the total Kraus operators can be described as $\mathbb{K}%
_{l}=\prod_{(i,j)}K_{q}^{(i,j)}$, with $q=1,2$ and the superscript $(i,j)$
going through all the component modes necessary to complete the state
transformation. Obviously, the class of $\left\{ \mathbb{K}_{l}\right\} $
still belongs to the SIO, and the index of the operators is $l=2,4,\ldots
,2^{n}$ ($n$ is the number of the steps in Fig.\thinspace 1). Therefore, we
can finally obtain $\sum_{l}^{2^{n}}\mathbb{K}_{l}|\psi \rangle \langle \psi
|\mathbb{K}_{l}^{\dagger }=|\phi \rangle \langle \phi |$.

\section{State transformation from 3D state into 2D state}

As an example, we consider the conversion $\psi _{1}|0\rangle +\psi
_{2}|1\rangle +\psi _{3}|2\rangle \rightarrow \phi _{1}|0\rangle +\phi
_{2}|1\rangle $ with the help of the ancillary modes. The devices in
Fig.\thinspace 3(b) realize the map as follows%
\begin{eqnarray}
|0V\rangle &\rightarrow &\cos 2\theta _{1}(\cos 2\theta _{4}|0\widetilde{0}%
H\rangle \!-\!\sin 2\theta _{4}|0\widetilde{0}V\rangle )\!+\!\sin 2\theta
_{1}|1\widetilde{1}H\rangle ,  \notag \\
|1V\rangle &\rightarrow &\cos 2\theta _{2}(\cos 2\theta _{5}|0\widetilde{1}%
H\rangle \!-\!\sin 2\theta _{5}|0\widetilde{1}V\rangle )+\sin 2\theta _{2}|1%
\widetilde{0}H\rangle ,  \notag \\
|2V\rangle &\rightarrow &\cos 2\theta _{3}|1\widetilde{0}V\rangle \!-\!\sin
2\theta _{3}|1\widetilde{1}V\rangle ,
\end{eqnarray}%
where, $\widetilde{0}$ and $\widetilde{1}$ distinguish the two groups of the
spatial modes split by the PBS, and\ both with the polarization modes $%
\left\vert H\right\rangle $ and $\left\vert V\right\rangle $\ acting as the
ancillary modes.\ Then the Kraus operators can be derived from the above map,%
\begin{eqnarray}
K_{1} &=&-\cos 2\theta _{1}\sin 2\theta _{4}|0\rangle \langle 0|+\cos
2\theta _{3}|1\rangle \langle 2|,  \notag \\
K_{2} &=&\cos 2\theta _{1}\cos 2\theta _{4}|0\rangle \langle 0|+\sin 2\theta
_{2}|1\rangle \langle 1|,  \notag \\
K_{3} &=&-\cos 2\theta _{2}\sin 2\theta _{5}|0\rangle \langle 1|-\sin
2\theta _{3}|1\rangle \langle 2|,  \notag \\
K_{4} &=&\sin 2\theta _{1}|1\rangle \langle 0|+\cos 2\theta _{2}\cos 2\theta
_{5}|0\rangle \langle 1|.  \label{Krause-3}
\end{eqnarray}%
These Kraus operators provide a general conversion process from 3D states
into 2D states. In the experiment, we consider a special case, i.e., the
input state is
\begin{equation}
|\psi ^{3}\rangle =\sqrt{\alpha }|2\rangle +\sqrt{(1-\alpha )/2}(|0\rangle
+|1\rangle ),
\end{equation}
with$~\alpha \in \lbrack 0,1/2]$ and the target state is $|\phi \rangle =%
\frac{\sqrt{2}}{2}(|0\rangle +|1\rangle )$. The\ angles of the HWPs in
Fig.\thinspace 3(b) are set as
\begin{eqnarray}
2\theta _{1} &=&2\theta _{2}=\arccos \left[ \frac{1}{\sqrt{2(1-\alpha )}}%
\right] ,  \notag \\
2\theta _{3} &=&-\pi /4,  \notag \\
2\theta _{4} &=&2\theta _{5}=-\arccos (\sqrt{1-2\alpha }).
\end{eqnarray}%
The output state of the whole system is
\begin{equation}
\frac{\sqrt{2}}{2}(|0\rangle \!+\!|1\rangle )\left[ \sqrt{\frac{1\!-\!\alpha
}{2}}(|\widetilde{1}H\rangle \!+\!|\widetilde{0}H\rangle )\!+\!\sqrt{\alpha }%
(|\widetilde{0}V\rangle \!+\!|\widetilde{1}V\rangle )\right] ,
\end{equation}%
from which we will obtain the target state by performing spatial tomography
on the modes of $|0\rangle $ and $|1\rangle $. Note that the tomography has
been done on both of the two groups of spatial modes to provide a
deterministic transformation.

\section{Analytical results of the distillation fidelity}

For any pure state and the incoherent operation $\mathcal{O}\in \{\text{MIO,
DIO, SIO, IO}\}$, the fidelity can also be described by the $m$-distillation
norm\thinspace \cite{One-Shot Distillation-1}:
\begin{equation}
F_{\mathcal{O}}(\left\vert \psi \right\rangle ,\left\vert \Phi
_{m}\right\rangle )=\frac{1}{m}\Vert |\psi \rangle \Vert _{\lbrack m]}^{2},
\label{fidelity_norm}
\end{equation}%
where $||\psi \rangle \Vert _{\lbrack m]}$ is the $m$-distillation norm
\begin{equation}
||\psi \rangle \Vert _{\lbrack m]}=\min_{|\psi \rangle =|x\rangle +|y\rangle
}\Vert |x\rangle \Vert _{l_{1}}+\sqrt{m}\Vert |y\rangle \Vert _{l_{2}},
\end{equation}%
where $||\bullet ||_{l_{1}}$ and $||\bullet ||_{l_{3}}$ are the $l_{1}$ norm
and $l_{2}$ norm. \ For a $d$-dimensional pure state, the $m$-distillation
norm can be described as
\begin{equation}
\Vert |\psi \rangle \Vert _{\lbrack m]}=\Vert \psi _{1:m-k^{\star
}}^{\downarrow }\Vert _{l_{1}}+\sqrt{k^{\star }}\Vert \psi _{m-k^{\star
}+1:d}^{\downarrow }\Vert _{l_{2}},
\end{equation}%
where $\psi _{1:k}^{\downarrow }$ denotes the vector consisting of the $k$
largest (by magnitude) coefficients of $|\psi \rangle $, and analogously $%
\psi _{k+1:d}^{\downarrow }$ denotes the rest of the coefficients. The
special number of $k^{\star }$ is defined by
\begin{equation}
k^{\star }=\text{arg}\min_{1\leq k\leq m}(\Vert \psi _{m-k^{\star
}+1:d}^{\downarrow }\Vert _{l_{2}}^{2}/k).
\end{equation}

To consider the conversion from the input state
\begin{equation}
|\psi ^{3}\rangle =\sqrt{\alpha }|2\rangle +\sqrt{(1-\alpha )/2}(|0\rangle
+|1\rangle )
\end{equation}
into a 2D target state, for $0\leq \alpha \leq 1/2$, the distillation
fidelity can be easily verified to be $1$ by using the $m$-distillation norm
presentation. It implies that the state $|\psi ^{3}\rangle $ can be
successfully converted to the 2D\ maximal coherence state $\left\vert \Phi
_{2}\right\rangle =\sqrt{1/2}\left( \left\vert 0\right\rangle +\left\vert
1\right\rangle \right) $\ by choosing a proper incoherent operation. While,
for $1/2<\alpha \leq 1$, it can be calculated that%
\begin{eqnarray}
&~&\Vert \psi _{2-k+1:3}^{3\downarrow }\Vert _{l_{2}}^{2}/k=1-\alpha \text{,}%
~~\text{for }~~k=1\text{,}  \notag \\
&~&\Vert \psi _{2-k+1:3}^{3\downarrow }\Vert _{l_{2}}^{2}/k=1/2\text{,}~~~~~%
\text{for}~~k=2\text{.}
\end{eqnarray}%
Thus we have $k^{\star }=1$, and
\begin{eqnarray}
\Vert |\psi ^{3}\rangle \Vert _{\lbrack 2]} &=&\Vert \psi _{1:1}^{\downarrow
}\Vert _{l_{1}}+\Vert \psi _{1+1:3}^{\downarrow }\Vert _{l_{2}}  \notag \\
&=&\sqrt{\alpha }+\sqrt{1-\alpha }.
\end{eqnarray}%
Finally, the distillation fidelity becomes
\begin{eqnarray}
F_{\mathcal{O}}(|\psi ^{3}\rangle ,\left\vert \Phi _{2}\right\rangle ) &=&%
\frac{1}{2}\Vert |\psi ^{3}\rangle \Vert _{\lbrack 2]}^{2}  \notag \\
&=&\frac{1}{2}(\sqrt{\alpha }+\sqrt{1-\alpha })^{2}.
\end{eqnarray}%
Obviously, a reasonable target state is $\left\vert \phi _{3\rightarrow
2}\right\rangle =\sqrt{\alpha }|0\rangle +\sqrt{1-\alpha }|1\rangle $, which
can reach the distillation fidelity above.

In an analogous way, for the transformation from the input state $|\psi
^{4}\rangle $ in Eq.$\,$(\ref{4D-input}) into a 3D target state, we can
obtain the fidelity
\begin{eqnarray}
F_{\mathcal{O}}(|\psi ^{4}\rangle ,\left\vert \Phi _{3}\right\rangle ) &=&%
\frac{1}{3}\Vert |\psi ^{4}\rangle \Vert _{\lbrack 3]}^{2}  \notag \\
&=&\left[ \sqrt{\frac{2\alpha }{3}}+\sqrt{\frac{2-4\alpha }{3}}\right] ^{2}
\end{eqnarray}%
for $\alpha \in \lbrack 0,1/6]\cup \lbrack 1/3,1/2]$. Thus a possible target
state is
\begin{equation}
\left\vert \phi _{4\rightarrow 3}\right\rangle =\sqrt{2\alpha }|1\rangle
\!+\!\sqrt{(1/2-\alpha )}(|2\rangle +|3\rangle ).
\end{equation}
While, for $\alpha \in \lbrack 1/6,1/3]$, the optimal value $F_{\mathcal{O}%
}(|\psi ^{4}\rangle ,\left\vert \Phi _{3}\right\rangle )=1$ can be reached,
which means that in this region the achievable target state is the maximally
coherent state $\left\vert \Phi _{3}\right\rangle =\sqrt{1/3}\left(
\left\vert 0\right\rangle +\left\vert 1\right\rangle +\left\vert
2\right\rangle \right) $.

\end{document}